# A solid-solution approach for room-temperature bulk plasticity in $KTa_{1-x}Nb_xO_3$

Alexander Frisch[1], Jiawen Zhang[2], Martin Setvin[3], Xuping Wang[4], Wenjun Lu[2*], Xufei Fang[1*]

[1]Institute for Applied Materials, Karlsruhe Institute of Technology, 76131 Karlsruhe, Germany

[2]Department of Mechanical and Energy Engineering, Southern University of Science and Technology, 518055 Shenzhen, China

[3]Department of Surface and Plasma Science, Faculty of Mathematics and Physics, Charles University, 18000 Prague, Czech Republic

[4]Advanced Materials Institute, Qilu University of Technology (Shandong Academy of Sciences), 250014 Jinan, China

*"The Tao produced One; One produced Two; Two produced Three; Three produced Infinity."*

*Tao Te Ching, by Lao-tsu*

**Abstract:**

Dislocations are being engineered into perovskite oxides to harvest versatile functional properties. One major bottleneck, however, persists: perovskite oxides that can be engineered with dislocations, particularly via mechanical deformation at room temperature in bulk scale, have so far been limited to only three materials: $SrTiO_3$ (2001, Brunner *et al.*), $KNbO_3$ (2016, Mark *et al.*), and $KTaO_3$ (2024, Fang & Zhang *et al.*). Here, we propose a simple and effective approach by using solid solution to significantly extend the range of materials. We showcase $KTa_{1-x}Nb_xO_3$ ($0<x<1$) perovskite oxides for their bulk plasticity at room temperature by constructing a closed-loop validation workflow that includes crystal growth, Brinell indentation, bulk compression, and transmission electron microscopy characterization. Our findings are expected to unlock the materials toolbox for dislocation-tuned functionality of perovskite oxides.

**Keywords:** dislocation; perovskite oxide; solid solution; potassium tantalate niobtate; plasticity

## 1. Introduction

Perovskite oxides are widely used in electronic devices made of electroceramics, for instance, as dielectrics in multilayer ceramic capacitors or in piezoelectric actuators or sensors for the conversion of electrical and mechanical energy. To further enhance their functionality, dislocations (one dimensional line defects) have been proposed as a new perspective for electroceramics engineering. Recent proofs-of-concept include ferroelectric [1, 2] and photocatalytic [3, 4] properties. The improvement of these functional properties via dislocation engineering has been primarily attributed to the charged dislocations and large strain field around the dislocation cores [5, 6].

Nevertheless, due to the brittleness of ceramic materials, it remains an outstanding challenge to engineer dislocations into ceramics without inducing cracks. Among the various experimental approaches for dislocation engineering [6-12], mechanical deformation through room-temperature cyclic indentation and scratching proves to be an effective tool for generating high-density dislocations up to ~$10^{15}/m^2$ in large plastic zones at meso-/macroscale [13]. This method, however, has been limited to ceramics that exhibit room-temperature bulk plasticity, in other words, that have low lattice friction stress to dislocation motion [14, 15] to prevent premature cracking prior to the onset of plasticity.

For perovskite oxides, which are of particular interest here, there have been so far only three materials reported to exhibit bulk plasticity mediated by dislocations at room temperature, namely, strontium titanate ($SrTiO_3$, STO) in 2001 [16], potassium niobate in 2016 ($KNbO_3$, KNO) [17], and potassium tantalate in 2024 ($KTaO_3$, KTO) [18]. The question remains open: are there more perovskite oxides that can be plastically deformed in bulk at room temperature?

Here, we report a promising approach by using solid solution of the two previously reported materials, KTO and KNO. The resulting material system is potassium tantalate niobate ($KTa_{1-x}Nb_xO_3$, KTN; $0<x<1$) [19-22], in which the B-site in the center of the perovskite unit cell is occupied by either $Ta^{5+}$ or $Nb^{5+}$ in a statistical distribution, as depicted in **Figure 1A**. As both ions carry the same charge state, substituting them for each other does not influence the species and concentration of charges in KTN. Nevertheless, changes in the room-temperature stable phases are to be expected [23]. At room temperature, pure KTO is cubic while pure KNO is orthorhombic. The latter undergoes phase transitions to the rhombohedral phase upon cooling to 263 K, to the tetragonal phase upon heating to 498 K and finally to the cubic phase upon heating to 708 K [24]. The room-temperature stable phase in KTN is therefore dependent on the Nb concentration [20].

As for room-temperature mechanical deformation, the stresses required for the onset of plastic deformation in bulk samples, namely, the yield stress, of the two parent perovskite oxides differ widely: bulk KTO can plastically deform at ~270 MPa [18] while bulk KNO only requires ~60 MPa [17], both along the <001> axis. It is expected that the yield stress of the KTN system shall be located within the two ends, if no solid solution hardening is present (**Figure 1B**, dashed line).

In this rapid communication, we test our working hypothesis on KTN crystals with different Nb concentration by beginning with the meso-scale Brinell indentation [25] and scratching [13] tests, which have been proved as an effective method to indicate bulk plasticity. Then, to complete the testing loop, uniaxial bulk compression tests are performed on available KTN crystals to obtain the yield stress and the macroscale plastic strain. Finally, dislocations are visualized by transmission electron microscopy to close the validation loop.

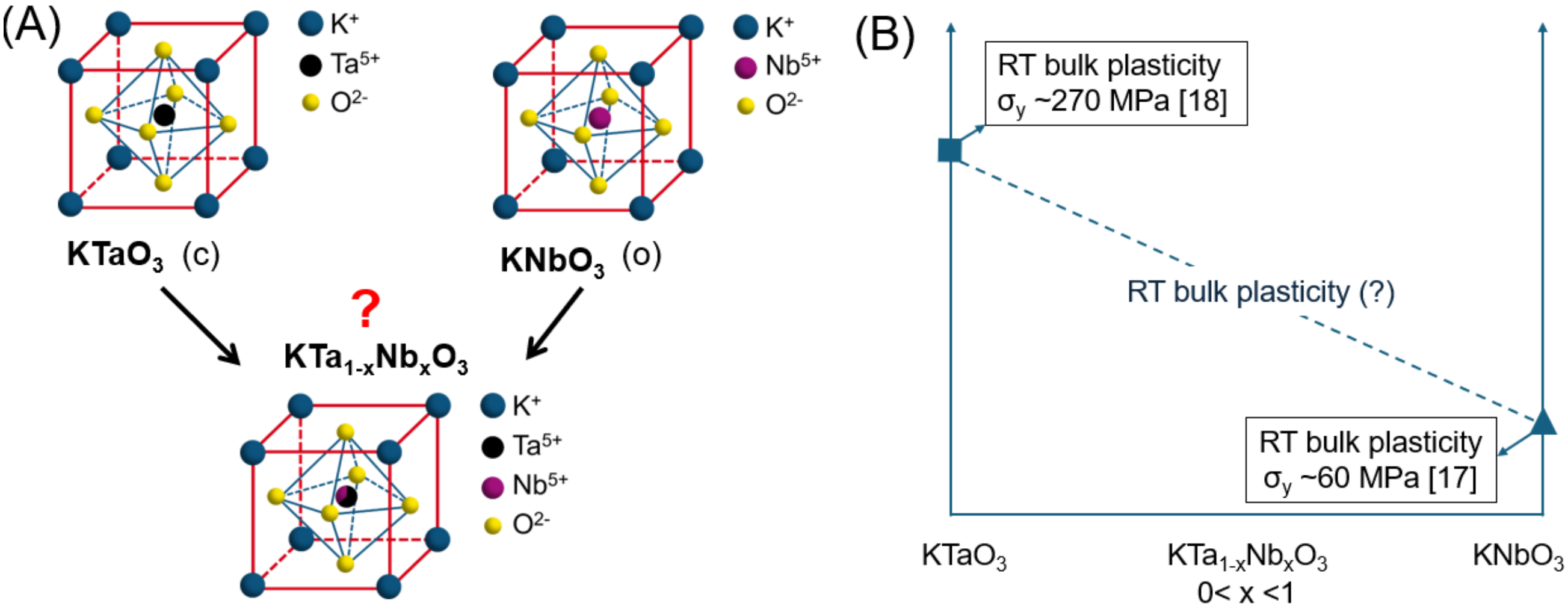

**Figure 1.** Concept for solid-solution based bulk plasticity in the $KTa_{1-x}Nb_xO_3$ system: the two parent-materials, $KTaO_3$ [18] and $KNbO_3$ [17], exhibit room-temperature bulk plasticity, as indicated in (B) by the two references and the yield stress, $\sigma_y$. Note that the dashed line in (B) is only hypothetical.

## 2. Materials and Methods

To avoid the complexity of grain boundaries on crack formation [26], $KTa_{1-x}Nb_xO_3$ single crystals were used for the experiments. For simplicity, we use KTNx to indicate the nominal Nb content (x). To screen a broader range of compositions as well as to test the reproducibility of the results from various providers, samples with x = 0.1, 0.175, and 0.3 were prepared by solidification from a melt (slow-cooling method [20]) and acquired from Oak Ridge National Laboratory. Samples with x = 0.3, 0.4, and 0.5 were grown by the Czochralski method by one of the current authors [27]. Prior to mechanical testing, the samples were first ground and then fine-polished to achieve mirror-surface finish. Note that below x ≈ 0.4 the cubic phase in the KTN solid solution is preserved at room temperature [20]. Surpassing this Nb content, tetragonal phase occurs at room temperature.

A universal hardness testing machine (Finotest, Karl-Frank GmbH, Weinheim-Birkenau, Germany), equipped with a hardened steel indenter sphere (diameter of 2.5 mm), was used to perform the cyclic Brinell indentation (**Figure 2A**) and Brinell scratching. For Brinell indentation, the hardened steel sphere was pressed into the sample under a load of 1.0 kgf, with a holding time of 10 s. For cyclic Brinell scratching, a load of 1 kgf was used on the sample, which is carried by a motorized one-axis stage (PI Line M-6832V4, Physik Instrumente GmbH & Co. KG, Karlsruhe, Germany) to create scratch tracks with a length of arbitrary length (e.g., 2 mm). More details regarding these two methods can be found elsewhere [13, 25]. Images of the indents and scratch tracks were captured using a 3D confocal LASER scanning microscope (LEXT OLS 4000, Olympus IMS, Waltham, USA).

For bulk compression tests, the samples were loaded along the <001> direction in a universal testing machine (MTS E45) at a constant strain rate of $\sim10^{-4}$ $s^{-1}$. We collected the deformation images using the VIC-gauge 2D software (Correlated Solution, Inc.). To mitigate the frictional stresses at the contact interfaces between the sample and the compression setup arising from the Poisson's effect, well-annealed copper foils of 0.5 mm in thickness were placed at both contact surfaces (between the sample and the alumina blocks) as a malleable interlayer, which is helpful in suppressing the sample fracture.

Dislocations are visualized and characterized using transmission electron microscopy (TEM). TEM lamellae were prepared using a focused ion beam (Helios Nanolab 600i, FEI) for lift-out inside the scratch tracks (dislocation-rich plastic zone). The TEM samples were then characterized at 200 kV in an TEM (FEI Talos F200X G2, Thermo Fisher Scientific, USA) with STEM mode. For ABF imaging, a probe semi-convergence angle of 17 mrad was employed, with inner and outer semi-collection angles of 13 and 21 mrad, respectively.

## 3. Results and Discussions

The indentation imprints on the samples surface are captured by laser confocal microscopy for several representative compositions (**Figure 2B-D**). The resulting crystallographic cross-hatched patterns in **Figure 2B-D** are the slip traces due to plastic deformation. The slip traces are arranged vertically and horizontally on the (001) surfaces, corresponding to the activated {110}<110> slip systems at room temperature, same for KTO [18] and KNO [17]. No domain patterns were observed under the polarized microscope for these two samples KTN0.175 and KTN0.3, as they remain cubic at room temperature. In contrast, KTN0.5 is of the tetragonal phase and shows a typical ferroelectric domain pattern (strips indicated by the white arrows in **Figure 2D**). Around the indentation imprint (white dashed circle), domain fragmentation around the slip traces (vertical and horizontal lines as in the case of KTN0.175 and KTN0.3) is clearly visible, showcasing the interaction between dislocations and ferroelectric domains in this crystal, closely resembling the deformation features as in pure KNO [28].

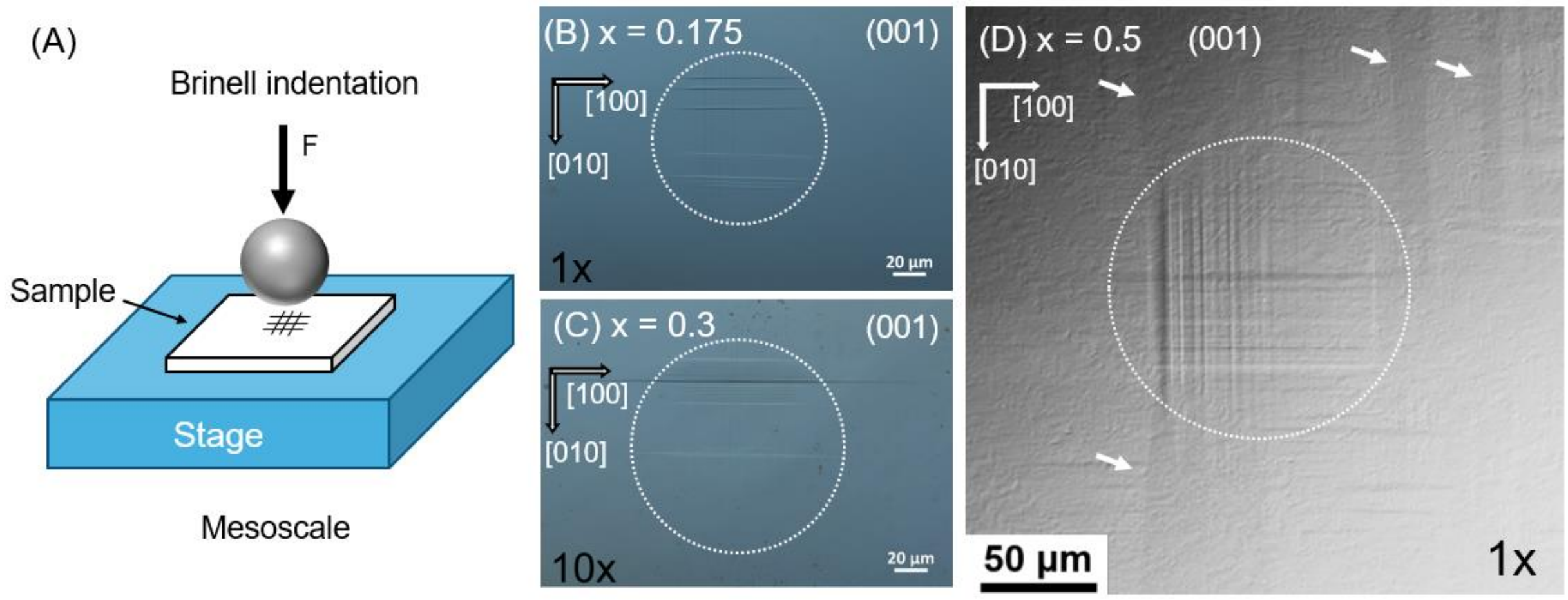

**Figure 2**. (A) Schematic illustration for Brinell indentation, where the indenter has a diameter of 2.5 mm in the current test. (B-D) Residual indent imprints (plastically deformed regions, or the plastic zones as indicated by the white dashed circles) of KTNx (B for x = 0.175, C for x = 0.3 from the slow-cooling method, and D for x = 0.5 as the representative samples), visualized by using laser confocal microscopy. Note the 1x and 10x in the figures indicate the number of cycles during indentation. Note for KTN0.5 the crystallographic notion is for pseudo-cubic.

Recently, the current authors have established a workflow on single-crystal $KTaO_3$ [18] by using the materials' plastic deformation response to Brinell indentation as a fast and effective probe to evaluate the feasibility of bulk plasticity in uniaxial compression. In short, through Hertzian contact mechanics, the maximum applied shear stress $\tau_m$ during indentation can be made by [29]:

$$\tau_m \approx 0.46 \frac{P}{\pi a^2} \tag{1}$$

where *P* is the applied load, *a* is the contact radius (for elastic loading). Take the composition of x = 0.3 (KTN0.3) as an example: consider that the post-mortem plastic zone (**Figure 2C**) has a diameter of about 140 μm, and the load was *P* = 1 kgf (9.8 N), this yields an estimated $\tau_m$ ≈ 320 MPa underneath the indenter using Eq. (1).

Since the Hertzian contact theory applies for elastic deformation [30], the $\tau_m$ under the Brinell indenter is expected to be larger than $\tau_{CRSS}$ (critical resolved shear stress in bulk compression of single crystals). It is therefore a reasonable deduction that the lattice friction stress for dislocation glide in KTN0.3 at room temperature shall be below 320 MPa, which serves as an upper bound estimation. This is indeed the case as validated by the bulk compression of KTN0.3 in **Figure 3**. Due to the loading axis being <001> and the Schmid factor of 0.5, the $\tau_{CRSS}$ will be half of the yield stress (~300 MPa from **Figure 3A**), taking the value of ~150 MPa.

Interestingly, the yield stress of ~ 300 MPa for KTN0.3 exceeds both the yield stress of ~270 MPa for pure KTO and ~60 MPa for pure KNO (as hypothesized in **Figure 1B**). It may be tempting to attribute this to the solid solution hardening, a usually observed strengthening effect for alloying in metals. This higher value for the yield stress in KTN0.3, however, needs to be taken with caution. First, the ionic radii of $Ta^{5+}$ and $Nb^{5+}$ in perovskite oxides are both 0.78 Å [31]. Normally a large difference between the ionic radii would cause lattice distortion to hinder the motion of dislocations, therefore increasing the yield stress [32]. As the ionic radii for $Ta^{5+}$ and $Nb^{5+}$ do not differ, negligible lattice distortion is expected to contribute to solid solution hardening. Second, the KTN0.3 sample prior to deformation clearly exhibits numerous horizonal strip features (**Figure 3B**, corresponding to point 1 in **Figure 3A**). It is unclear at this stage what these strips are, but given the fact that these strips persist during the following-up deformation process, they are immobile and not ferroelectric domains, but most likely some structural defects induced during the crystal growth, which remains a common challenge in growing KTNO crystals [33]. These structural defects can effectively block dislocation motion to increase the yield stress. Nevertheless, the absolute values for the yield stress for the various compositions as well as for the pure crystals of KTO and KNO is not of major concern here (although it may seem to violate the hypothesis in **Figure 1B**), as these values can also vary significantly depending on the defect chemistry, impurity level, and pre-existing dislocation densities, with a typical example being STO for recent systematic studies [34-36]. More of interest is the overall range of these yield stress values, which indicate the lattice friction stresses that are much lower than most of the structural ceramics (exceeding several GPa) [11]. These relatively low lattice friction stresses (lower than the fracture strength) are desirable for room-temperature bulk plasticity in KTN as well as in KTO [18], KNO [17], and STO [16], as dislocation glide and multiplication are facilitated prior to the fracture of the samples.

In **Figure 3A**, we also note the abnormal initial nonlinear segment ("softening" as highlighted by the rectangle) in the stress-strain curve. This is caused by the copper foil placed between the sample and the loading cell to avoid cracking of the sample (see **Section 2** for details). The annealed copper foil is very soft and acts as a buffer layer to mitigate potential stress concentration at the sharp edges of the sample. This buffer layer results in a shift of the absolute plastic strain value but does not affect the yield stress value according to the Maxwell model. As the plastic deformation proceeds (**Figure 3C, D**), slip traces that are 45°-inclined to the <001> loading axis appear, again confirming the {110} slip planes are activated. Further loading will eventually lead to crack formation (**Figure 3E**).

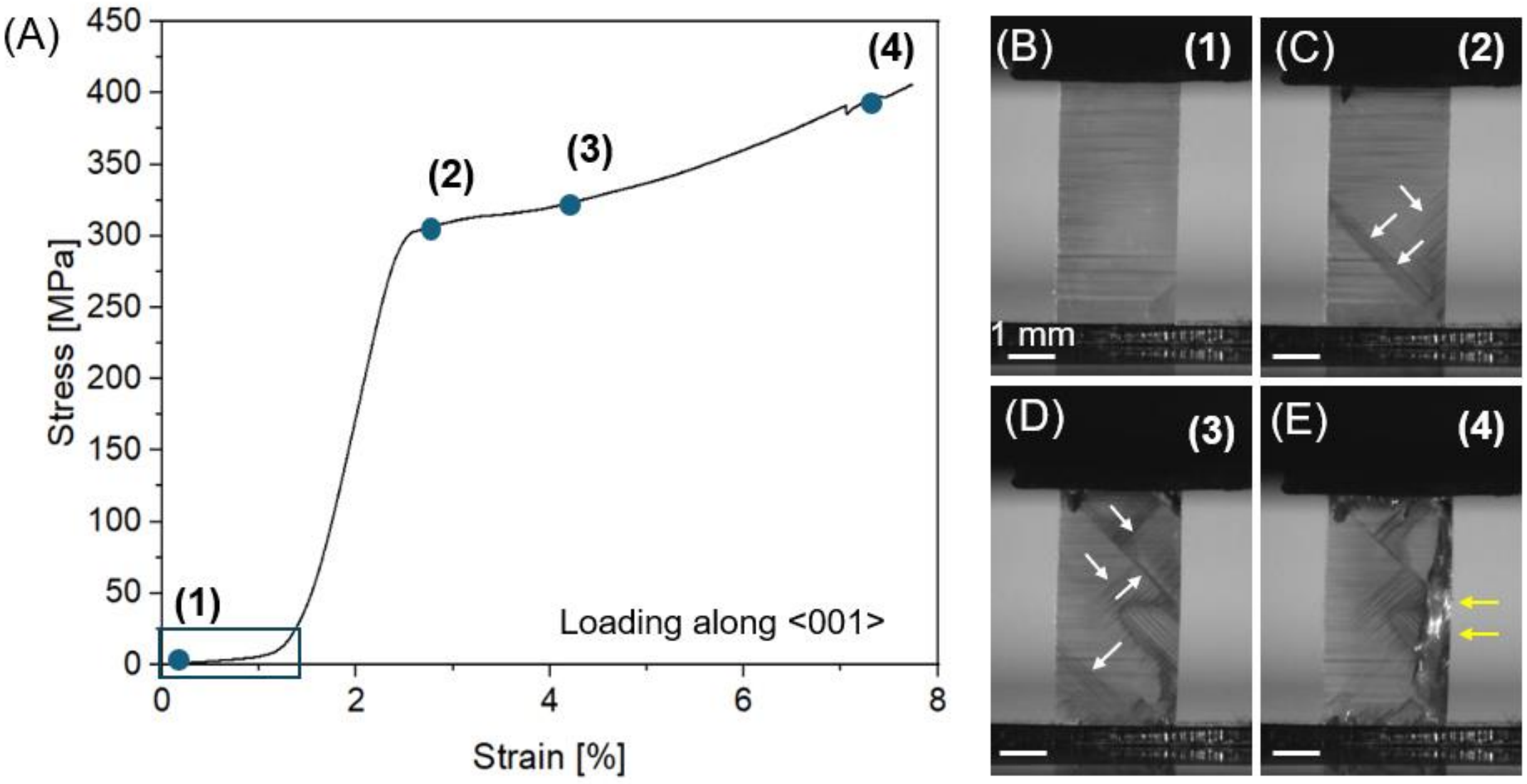

**Figure 3.** Bulk compression of KTN0.3 (from Czochralski method) at room temperature: (A) engineering stress-strain curves; (B-E) Side-view snapshots of the sample's deformation process. The white arrows in (B, C, D) indicate the slip traces and the yellow arrows in (E) indicate the cracks (white contrast). Loading axis is <001>, and the sample is placed between the two alumina plates (dark rectangles at the top and bottom).

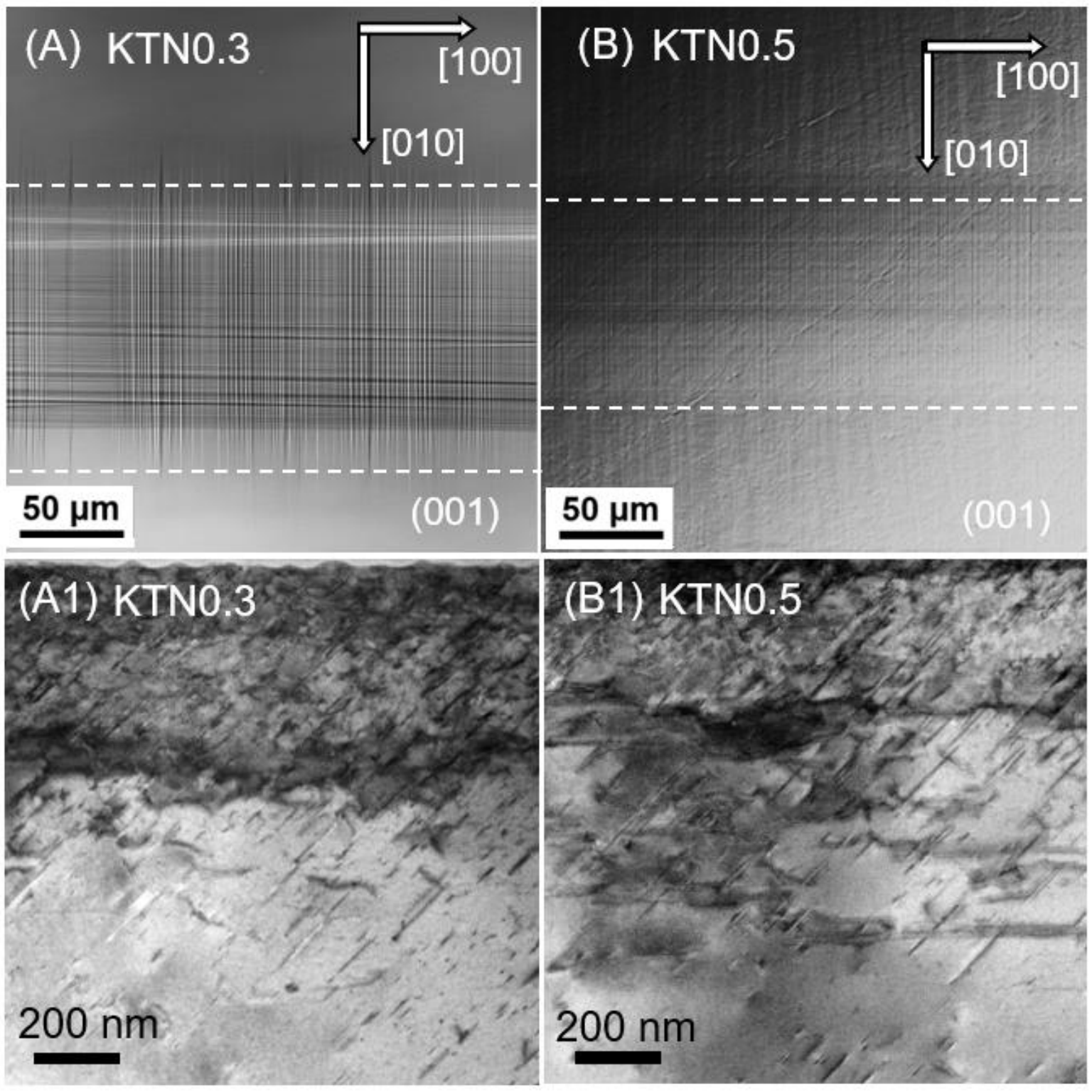

**Figure 4.** Brinell scratching tests on (A) KTN0.3 (cubic) and (B) KTN0.5 (tetragonal) showcasing the mesoscale plastic zone on the sample surface. 10-pass (10x) scratching was performed to generate high-density dislocations. Images (A, B) are obtained on a laser microscope. Corresponding ABF-STEM images for the dislocation structures for (A1) KTN0.3 and (B1) KTN0.5.

To complete the validation loop, we further carried out ABF-STEM analysis to examine the dislocation structure. To this end, we used cyclic Brinell scratching (10-pass, or 10x scratching) to generate high-density dislocations in both KTN0.3 (**Figure 4A**) and KTN0.5 (**Figure 4B**) in a large plastic zone at mesoscale. The images in **Figure 4A1, B1** show direct evidence of the high-density dislocations inside the scratch tracks. The dislocation density after 10x scratching is estimated to be around ~$10^{14}/m^2$, similar to the case of $KTaO_3$ [18].

## 5. Conclusion

In this work, we demonstrate that solid-solution perovskite oxides, potassium tantalate niobate, with a Nb fraction (atomic) up to 0.5 can be plastically deformed in bulk at room temperature. The results suggest an infinite possibility of the composition $KTa_{1-x}Nb_xO_3$ ($0<x<1$) shall possess room-temperature bulk plasticity, provided that the crystals can be synthesized, as no classic solid solution hardening was observed. These findings open the path for active discovery of more plastically deformable oxides via the solid-solution approach, and pave the road towards dislocation-engineered functional ceramics besides the previously reported but limited number of perovskite oxides ($SrTiO_3$, $KNbO_3$, and $KTaO_3$).

**Acknowledgement:**

A. Frisch and X. Fang acknowledge financial support by the European Union (ERC Starting Grant, Project MERCERDIS, grant No. 101076167). Views and opinions expressed are, however, those of the authors only and do not necessarily reflect those of the European Union or European Research Council. Neither the European Union nor the granting authority can be held responsible for them. X. Wang acknowledges the Science-Education-Industry Integration Innovation Pilot Project of Qilu University of Technology (grant No. 2026GH35). M. Setvin and X. Fang would like to thank L. A. Boatner at Oak Ridge National Laboratory for providing some of the KTNO crystal for the indentation tests. W. Lu has received the following grants for supporting this work: Shenzhen Science and Technology Program (grant No. JCYJ20230807093416034), the Open Fund of the Microscopy Science and Technology-Songshan Lake Science City (grant No. 202401204); National Natural Science Foundation of China (grant No. 52371110). We thank O. Preuß for the helpful suggestions on the bulk compression test, and we also acknowledge the use of the facilities at the Southern University of Science and Technology Core Research Facility.

**Data Availability:**

All relevant data are available in the main text.

**Competing Interests:**

The authors declare no competing interests.

**Declaration of AI usage:**

The authors declare no use of AI generated content (neither figure nor text) in preparing this manuscript.

**Author contribution:**

X.F.: conceptualization, resources, supervision, writing, revision; W. L.: resources, supervision, revision; A.F.: lead in experiment, data analysis, writing; J. Z.: support in experiment, data analysis, revision; M. S.: resources, revision; X. W.: resources, revision.